\documentclass[pdflatex,a4paper,english,11pt,oneside,ja=standard]{bxjsarticle}

\usepackage[T1]{fontenc}
\usepackage{amsmath,amssymb,amsfonts}
\usepackage[pdftex]{graphicx}
\usepackage{xcolor}

\usepackage{comment} 
\usepackage{siunitx} 
\usepackage{booktabs} 
\usepackage{array}    
\usepackage{arydshln} 
\usepackage{makecell} 

\usepackage[hyphens]{url}
\usepackage[hidelinks]{hyperref}

\usepackage{latexsym}
\usepackage{multirow} 
\usepackage{threeparttable} 

\usepackage[hang,small,bf]{caption}
\usepackage[subrefformat=parens]{subcaption}
\usepackage{pdflscape}

\usepackage[backend=biber, style=numeric, sorting=none]{biblatex}
\title{
DLT Terms 100\\: a learner's glossary based on ISO 22739:2024
}

\author{
  \begin{tabular}[t]{c}   Yusuke KANEKO
  \thanks{ 
    Corresponding author. 
    E-mail: \texttt{kaneko.yusuke@jri.co.jp}
  } \end{tabular}
  \hfill
  \begin{tabular}[t]{c} Takumi AITA \end{tabular} 
  \\[0.5em]
  \begin{tabular}[t]{c} Kohei ICHIHARA \end{tabular}
  \hfill
  \begin{tabular}[t]{c} Hiroki WATANABE \end{tabular}
  \\[1em]
  \small The Japan Research Institute, Ltd. \\
  \small Shinagawa, Tokyo 141--0022 Japan
}

\date{}

\begin{document}
\maketitle

\begin{abstract}
We propose ``DLT Terms 100'' aimed at improving literacy regarding blockchain (distributed ledger technology: DLT).
The pace of DLT's social implementation has been slower than initially expected, partly due to a shortage of engineers and literate individuals.
In particular, the discrepancy between business and technical terminology, and the confusion caused by the repurposing of existing financial terms, are barriers to understanding the concepts and technologies.
Our proposed glossary is based on ISO 22739:2024, and categorizes and labels terms based on their frequency of appearance in representative domestic and international literature and their difficulty level.
\end{abstract}

\section{Introduction \& Summary}
\subsection{Motivation}
Blockchain and Distributed Ledger Technology (DLT) have advanced through proof-of-concept (PoC) trials \cite{kaneko_DeExam, Aita_ISTAS23, Ichihara_ICCE2024} and practical applications in specific fields, and global standardization efforts \cite{Chen2022, Qi2024}. 
However, widespread social implementation has not yet been achieved compared to early expectations. 
This is partly due to a shortage of highly skilled engineers and a lack of literacy to evaluate the technology correctly. 
While companies conduct employee training to improve DLT literacy, the rising cost of education makes establishing efficient programs an urgent priority. 
In Japan, ``blockchain'' was added to the IT Passport exam syllabus in 2019 \cite{ipa2018_ip_syllabus40_e}, making DLT a required basic knowledge for professionals.

A primary barrier to understanding DLT is the terminology gap between business practitioners and engineers, arising from two factors. 
First, confusion caused by financial analogies; for example, while a ``wallet'' manages private keys, many mistakenly believe assets or non-fungible tokens (NFTs) are stored inside it \cite{Ichihara_2023}. 
Second, interdisciplinary complexity; a ``token'' refers to assets in DLT but an authentication device in security. 
These overlaps delay comprehension for both business and IT departments. 

An effective solution to these challenges is to clearly define terms and their meanings. 
In the early stages of DLT, it was often explained that ``blockchain'' was an undefined term, but ISO 22739 was released in 2020. Based on this, the Japanese Industrial Standard (JIS X 22739) was released in February 2026.

\subsection{Purpose \& Summary}
Our research question is: What technical commentary materials effectively improve DLT literacy?
We propose the ``DLT Terms 100'' (Table \ref{tab:DLT100}), a learner's glossary based on ISO 22739:2024 \cite{ISO22739}, balancing international standards with practical application. 

Currently, empirical experiments and evaluations to verify the effectiveness of our proposed glossary have not yet been conducted. 
Therefore, future work will evaluate how much this list reduces communication costs between engineers and practitioners.

\section{Proposal of ``DLT Terms 100''}
The ``DLT Terms 100'' we propose is designed to serve as a guide for beginners to progressively deepen their learning. 
To achieve this, the 100 terms included in ISO 22739 were categorized into eight domains, and the terms in each category were organized into three levels of difficulty.

\subsection{Domains}
DLT involves multiple closely related specialized domains, ranging from fundamental IT system components to economic incentives related to DLT. 
On the other hand, it involves a high learning cost and may not always be necessary for all practitioners and engineers to cover all domains. 
Based on this reality, the included terms were categorized into the eight domains shown in Table \ref{tab:DLT100}.

\subsection{Classification by Difficulty}
The difficulty of each term was classified into three levels (1 to 3) based on the following criteria. 
The difficulty levels were set based on the inclusion of the terms in the in-house training materials used at the author's company, as well as the comprehension level of the users of these materials, establishing the importance of the terms for practitioners. 
It is recommended that all practitioners understand low-difficulty terms and that DLT specialists understand and acquire High-difficulty terms selectively depending on their work field.

\textbf{Difficulty 1:} Basic terms that serve as an entry point to the technology, appearing in news and general business books. 
Even non-experts have heard of them, and their meanings can be inferred from the context.

\textbf{Difficulty 2:} Practical terms related to the specific operations of DLT and the processes required when using services. 
These terms are necessary for actually handling the technology or understanding intermediate-level explanatory articles.

\textbf{Difficulty 3:} Specialized terms that involve implementation-level details, distributed computing theory, and advanced governance. 
These terms are often strictly handled in official documents and academic papers.

\subsection{Term Highlighting via Frequency Analysis}
\label{sec:ISO22739_survey}

We investigated the extent to which the terms included in ISO 22739 align with commercially available technical references.
Specifically, we checked the presence or absence of the 100 headwords included in ISO 22739:2024 and their synonyms in the following four references: 
1) ISO 23257:2022 (Blockchain and distributed ledger technologies — Reference architecture) \cite{ISO23257},
2) \textit{Mastering Bitcoin}, 3rd Edition \cite{Mastering_Bitcoin},
3) \textit{Mastering Ethereum}, 2nd Edition \cite{Mastering_Ethereum},
4) \textit{Introduction to Blockchain Technology} (in Japanese) \cite{yamasaki2021_e}.

As a result of the investigation, the coverage rates of the terms were 77\% for 1), 59\% for 2), 67\% for 3), and 71\% for 4).
There were 37 terms that appeared in all four investigated documents, 71 terms that appeared in two or more documents, and 5 terms that did not appear in any of the four investigated documents.

The terms (23 terms) that do not appear in ISO 23257 can be considered of low importance.
On the other hand, the terms (5 terms) presented at the beginning of ISO 23257 can be considered important for using the document, and it cannot be assumed that the reader has already understood them.
Furthermore, the terms (15 terms) that appeared only in ISO 23257, all of which are compound words with the prefix DLT such as ``DLT account'', can be said to be not yet widely used.

\subsection{Limitations and Direction of Revision}
This glossary does not include trademarks in accordance with the fact that ISO 22739:2026 does not include specific product names or service names (trademarks, etc.).
In practice, it is also necessary to understand DLT-related trademarks, so it may be considered to add trademarks, etc., as appropriate when using the list.

\bigskip

\begin{landscape}
\begin{table}[h] 
  \centering
  \caption{DLT terms 100.
  Synonyms are in parentheses. 
  DLT may be written as ``distributed ledger technology''.\\
  Legend: \textbf{Bold} indicates presence in all references in \cite{ISO23257, Mastering_Bitcoin, Mastering_Ethereum, yamasaki2021_e} (37 terms); 
  * omitted from ISO 23257:2022 (Use Cases) (23 terms); 
  ** absent from all references in \cite{ISO23257, Mastering_Bitcoin, Mastering_Ethereum, yamasaki2021_e} (5 terms); 
  $\dag$ defined in other ISO standards (12 terms); 
  $\ddag$ listed in the introductory section of ISO 23257:2022 (5 terms).}
  \label{tab:DLT100}
  \sffamily
  \begin{tabular}
    {p{25mm}|p{52mm}|p{64mm}|p{52mm}}
    \hline
    Category & Difficulty 1 (Basic) & Difficulty 2 (Practical) & Difficulty 3 (Specialized) \\
    \hline\hline

    1) System Infra. \& Components & 
    \textbf{blockchain}, blockchain system, blockchain technology*, distributed ledger, 
    DLT (distributed ledger technology), \textbf{peer-to-peer}
    & 
    decentralized system, 
    \textbf{distributed system}, 
    DLT solution, 
    DLT system (distributed ledger system), 
    DLT node, \textbf{node}
    & 
    DLT network, 
    DLT platform, \textbf{entity} \\
    \hline

    2) Data Structure \& Recording & 
    \textbf{block}, \textbf{ledger}, \textbf{record}$^{\dag}$ & 
    \textbf{block data}, 
    ledger record, 
    shared ledger*, 
    \textbf{timestamp}$^{\dag}$ 
    & 
    block header*, 
    \textbf{genesis block}, 
    hash link (cryptographic link)**, 
    leaf node*, 
    Merkle root*, 
    Merkle tree*, 
    orphan block*, 
    prune,     
    root node** \\
    \hline

    3) Cryptography \& Security & 
    \textbf{digital signature}$^{\dag}$, 
    \textbf{cryptography}$^{\dag}$, 
    immutability,
    \textbf{private key}$^{\dag}$, 
    \textbf{public key}$^{\dag}$ &  
    \textbf{hash function} (cryptographic hash function)$^{\dag}$, 
    hash value*$^{\dag}$, 
    public key cryptography*
    & 
    \textbf{identifier}, \textbf{nonce} \\
    \hline

    4) Consensus \& Validation & 
    \textbf{failure}$^{\dag}$ & 
    \textbf{confirmed}, 
    confirmed block, 
    confirmed transaction,   
    \textbf{consensus}$^{\ddag}$, 
    consensus mechanism, 
    \textbf{miner}, 
    \textbf{mining},     
    \textbf{validation}, 
    validator
    & 
    fault tolerance$^{\dag}$, 
    finality$^{\ddag}$, 
    \textbf{validated} \\
    \hline

    5) Transactions \& Operations & 
    & 
    DLT account, 
    off-chain*, off-ledger, 
    \textbf{on-chain}, on-ledger, 
    \textbf{transaction}, 
    \textbf{transaction fee},
    transaction record 
    & 
    DLT address**, 
    double spending*, 
    \textbf{interoperability}$^{\dag}$ \\
    \hline

    6) Assets \& Incentives & 
    \textbf{asset}$^{\dag}$ , 
    cryptoasset*, 
    \textbf{cryptocurrency},
    digital asset, 
    \textbf{token}, 
    \textbf{wallet} & 
    fungible token*, 
    non-fungible token (NFT)*, 
    reward system (incentive mechanism),
    security token* & 
    block reward, 
    fungible$^{\ddag}$, 
    non-fungible, 
    utility token* \\
    \hline

    7) Governance \& Access Control & 
    DLT user & 
    \textbf{hard fork}, 
    ledger split (fork),  
    \textbf{permissioned}, 
    permissioned DLT system, 
    permissionless, 
    permissionless DLT system, 
    private DLT system, 
    public DLT system, 
    soft fork
    & 
    DLT governance$^{\ddag}$ \\
    \hline

    8) Applied Tech, Identity \& Scalability & 
    \textbf{identity} & 
    decentralized application (Dapp)* ,
    \textbf{smart contract}$^{\ddag}$ & 
    decentralized identifier (DID)*, 
    decentralized identity, 
    DLT bridge**, 
    DLT oracle, 
    self-sovereign identity (SSI)**, 
    sidechain, 
    subchain \\
    \hline
  \end{tabular}
\end{table}

\end{landscape}

\clearpage
\printbibliography[title=References]

\end{document}